# Shades of Perception: User Factors In Identifying Password Strength

Jason M. Pittman[1], Nikki Robinson

[1] *Department of Computer Science, High Point University, High Point NC*

**Abstract**
**Purpose -** The purpose of this study was to measure whether participant education, profession, and technical skill level exhibited a relationship with identification of password strength.
**Design/methodology/approach** – Participants reviewed 50 passwords and labeled each as *weak* or *strong*. A Chi-square test of independence was used to measure relationships between education, profession, technical skill level relative to the frequency of weak and strong password identification.
**Findings -** The results demonstrate significant relationships across all variable combinations except for *technical skill* and *strong* passwords which demonstrated no relationship.
**Research limitations/implications** - This research has three limitations. Data collection was dependent upon participant self-reporting and has limited externalized power. Further, the instrument was constructed under the assumption that all participants could read English and understood the concept of password strength. Finally, we did not control for external tool use (i.e., password strength meter).
**Practical implications** - The results build upon existing literature insofar as the outcomes add to the collective understanding of user perception of passwords in specific and authentication in general. Whereas prior research has explored similar areas, such work has done so by having participants create passwords. This work measures perception of pregenerated passwords. The results demonstrate a need for further investigation into *why* users continue to rely on weak passwords.
**Originality/value** - The originality of this work rests in soliciting a broad spectrum of participants and measuring potential correlations between participant education, profession, and technical skill level.

**Introduction**
Password-based authentication is a prominent feature in modern life. Logging into a computer or application with a password is perhaps the most common event occurring in daily life. It goes without saying that much of the security for such systems rests upon passwords. Meanwhile, cyberattacks are targeting password authentication as a susceptible link in the defense chain. Nefarious actors are finding success with phishing, social engineering, and outright system attacks. Unfortunately, two factor authentication methods are not widely used (Mao, Florencio, and Herley, 2012). To upgrade to a multi factor authentication system, an organization would need to undertake a large-scale engineering effort and completely alter the way users interact with systems (Mao, Florencio, and Herley, 2012). As the undertaking of creating a two-factor authentication is too great for most organizations, or unavailable to public websites, users must continue to use password-based login methods.

Yet, password authentication has grown on such a scale that users are overburdened cognitively (De Joode, 2012). Existing research demonstrates that users maintain approximately 25 total password-protected accounts (Florencio and Herley, 2007; Dhamija and Dusseault, 2008). The same research also revealed that users enter a password eight times, on average each day. Overall, conventional text-based passwords are insecure (Florencio and Herley, 2007). The reasons for this have been well studied and there has been great effort to combat the seemingly inherent flaws in conventional password authentication through variations in form and recall modality (Jermyn et al. 1999; Brostoff and Sasse, 2000; Jansen, 2003; Wiedenbeck et al., 2005).

Concurrent to the development of alternatives to conventional password schemes, researchers have turned to investigating means to increase the strength of text-based passwords. Ur et al. (2012) studied 2,931 password creation subjects using 14 different password meters. Their findings (2012) suggested more stringent meters led users to create stronger passwords, as well as the inclusion of particular visual appearances. Of course, developing alternatives requires understanding as many facets of the underlying issues as possible. Accordingly, Ur et al. (2015) asked users to create passwords in an attempt to uncover potential patterns in password generation. The researchers also interviewed these users and found pervasive misconceptions associated with what password creation factors lead to *strong* or *weak* passwords. However, we feel the literature presupposes that general password users can meaningfully identify such.

Such misconceptions appear to be prevalent in password meter development communities as well. Carnavalet and Mannan (2014) found, "highly inconsistent strength outcomes for the same password in different meters, along with examples of many weak passwords being labeled as strong or even excellent" (pg. iii). However, the existing literature presupposes that users can identify password strength. As Carnavalet and Mannan noted, millions of users use password-strength checkers on web services that require users to create passwords but the password checkers are only partially effective in aiding users to create stronger passwords. Specifically, Carnavalet and Mannan (2014) determined inconsistencies in the password checkers may lead users to misunderstand the characteristics required in a stronger password.

Based on Ur et al. (2015), it can be inferred that users hold misguided ideas regarding password strength in the context of creating passwords. This led us to wonder if an underlying motivation for such misunderstanding might be related to the users themselves. That is, apart from the inconsistency in password checkers, perhaps users interact with passwords inconsistently. Of course, all such speculation presupposes users can differentiate between weak and strong passwords to begin with. Following on their prior study, Ur et al. (2016) demonstrated users indeed can reliably judge password strength despite not necessarily creating or using strong passwords themselves.

Such as disconnect between *knowing* and *action* caught our interest. Accordingly, we wondered what socioeconomic characteristics, if any, led participants to identify weak and strong password strengths in a statistically significant manner. Thus, we asked users to rate 50 passwords as either *strong* or *weak* as a means to develop more understanding of how individuals perceive passwords. At the same time, we gathered level of education, profession, password usage and management, electronic device usage, and self-assessed technology skills, as socioeconomic characteristics of interest.

**Background**

Characteristics of weak versus strong passwords vary across research even within the same time period. This is further confused by variation between password strength visualization tools such as found by Salem, Hossain, and Kamala (2008) and Zhang-Kennedy, Chiasson, and Biddle (2011). Salem, Hossain, and Kamala (2008) investigated how individuals cannot identify a strong password, and further, the word *strong* is not well defined in context to password strength. Salem, Hossain, and Kamala (2008) then created a tool to identify password strength based on well known password cracking techniques; dictionary attacks, time to crack, and shoulder surfing. Zhang-Kennedy, Chiasson, and Biddle (2011) focused on visualization of password cracking techniques through posters and educational comics to train individuals to create strong and memorable passwords. The focus of Zhang-Kennedy, Chiasson, and Biddle's (2011) study was to train users for a week and find if this improved knowledge to create stronger passwords to combat targeted, dictionary, and brute-force attacks.

Using existing literature, we established the following definitions for weak and strong passwords. A weak password is (a) a string of less than seven characters; (b) a string comprised of only alphabetical or numerical characters; (c) a string, when of mixed characters, has the numerical characters at the end of the string only; (d) a string, when of greater than seven characters or containing special characters, that is predictable (Yan, et al. 2000; Korkmaz and Dalkilic, 2010). Conversely, according to the same research, strong passwords (a) do not include any of the weak password characteristics; (b) include one or more numerical, punctuation, and uppercase character.

Notoatmodjo and Thomborson (2009) noted that users are the weakest link in controlling and maintaining password security. The reason being users are prone to create weak passwords and then reuse those potentially weak passwords (Notoatmodjo and Thomborson, 2009). The study done by Notoatmodjo and Thomborson (2009) found that almost half of the participants reused a password in an important account. Komanduri et al (2013) conducted their own study and found the same as Notoatmodjo and Thomborson (2009), users create weak passwords. To deal with this issue, researchers like Komanduri et al (2013) created telepathwords. These telepathwords were intended to discourage individuals from using predictable passwords (Komanduri et al, 2013).

Related to our study, Stavrou (2017) found that text-based passwords still create problems for users, even with the increased amount of password-strength checkers and password policies. Stavrou (2017) found that bad password construction practices can lead to weak passwords. While our study intended to find if users are able to identify weak or strong passwords, Stavrou (2017) created a conceptual architecture to assist users in creating a strong password.

In summary, knowledge of password strength appears to be a problem worth solving, but there has not been sufficient research into how individuals identify weak and strong passwords. There is a gap in the research that this study may be able to address; understanding of password strength based on an individual's background and experience with technology.

**Methodology**
Broadly, we speculated participants would be able to identify weak passwords consistently. Password characteristics such as length, capitalization, inclusion of alphanumeric and symbol characters likely serve as significant context clues. Further, we imagined participants would not be able to consistently identify strong passwords, particularly when such were intermingled with weak passwords of similar length and combination. More technically, the goal of this correlational research was to determine which socioeconomic characteristics, if any, have measurable interactions with password identification and to what extent any such correlation is positive or negative. To facilitate such, we operationalized *subject education level, profession, and self-reported technical skill* as socioeconomic variables on one hand and *successful identification of weak and strong passwords* as password identification variables on the other. Further, we imagined a single instrument as a means of collecting data to evaluate our hypotheses.

*Instrumentation*
We designed a single data collection instrument, portioned in three sections. The first section implemented a standard informed consent, including opt-out procedures, and required affirmation of participation before continuing to the second section. No personally identifiable information was collected, and data were coded to participant by a simple integer index ranging between 1 and 436. This section also indicated the study was being conducted with IRB approval and listed relevant IRB contact information should participants have questions or concerns.

Demographic questions were asked in the second section. We asked participants to self-report on age, gender, profession, technical skill level relative to others they knew, as well as how many passwords they used on a daily basis. The first three questions in this section served to collect categorical data for our socioeconomic variables. Further, we designed the last question as a screening mechanism insofar as we wanted to include only those individuals using at least one password daily.

Finally, the third section contained 50 passwords, each associated with a bounded response set of *weak* and *strong*. This facilitated collection of categorical data at the heart of participants' perceptions. Further, the binary nature of the responses directly aligned with current guidance for digital authentication via passwords (Grassi, Garcia, and Fenton, 2017). The 50 passwords were randomly generated in two phases according to standardized definitions of weak and strong (Carnavalet and Mannan, 2014; Grassi, Garcia, and Fenton, 2017) in terms of character length, character set composition, symbols and so forth. We developed a simple Python program to create the password lists, given the standard password strength definitions, for each phase as follows.

*Password construction procedure*
The first phase generated 100 weak passwords, parameterized as length of one-to-seven characters in length, the set of characters bounded to alphanumeric only, and any numerical characters positioned at the end of the string. Concurrently, the first phase generated 100 strong passwords, parameterized as greater than eight characters, the set of characters bounded to alphanumeric, punctuation, and special symbols, and the numeric or symbolic characters randomly placed within the string. The second phase consisted of removing any obvious weak (e.g., a single character or short, blatant sequence like *123*) and injection of five algorithmically weak passwords that may appear strong (e.g., G@m30f7hr0n3$). We then selected every second password in each strength category, yielding 25 passwords from each and 50 total.

*Participants*
To achieve a suitable sample, we started with a robust population framing (essentially, open-ended) of international *users*. Because of this, we elected to avoid traditional questionnaire instrumentation routes due to the tendency to return poorly sized subject pools. Instead, we opted to use Amazon's Mechanical Turk.

By using Mechanical Turk, we were able to solicit individuals from a general, cross-sectional population (scoped to Mechanical Turk users and biased towards the subset willing to participate in a questionnaire-based study) as opposed to a specific profession, age, or education category. According to Amazon, Mechanical Turk has a disparate and global user population of more than 500,000 people from over 190 countries. Further, research (Paolacci, Chandler, and Ipeirotis, 2010) has shown Mechanical Turk to be a strong and reliable platform for crowdsourcing participation across a wide spectrum of research modalities.

Participants earned .50 cents (USD) for completing our web based questionnaire of nine demographic data points consisting of: education level, technological skill level, profession, ability to perceive a password as either strong or weak, gender, age, number of passwords used on a daily basis, the use of a password manager, time per week using an electronic device and country in which they resided. Participants then viewed a series of passwords and were asked to rate them as either strong or weak.

*Sample Description*
The final sample size was 436 participants. Whereas the initial sample consisted of 447 individuals, we removed 11 participants from the data who did not complete the questionnaire entirely. We did not provide participants with any instructional information regarding password strength. While not controlled for, we did ask participants to avoid (a) searching for password strength definitions; (b) utilize tools such as password strength checkers; (c) or any form of outside help. That said, we recognize a limitation in our protocol exists insofar as we did not control for any of these behaviors.

The sample was demographically distributed across six categories when considering age (Figure 1). The majority of participants reported an age of between 25 and 34 years ($N$=205). The largest age group potentially speaks to the number of individuals who are familiar with technology in general and willing to participate in an online survey of this type. The next largest age range was 35-44 (102), which when combined with the 25-34 group, encompassed 300 of the total group. Only two participants were under the age of 18, making them members of a potentially vulnerable population. This will be addressed in the section below.

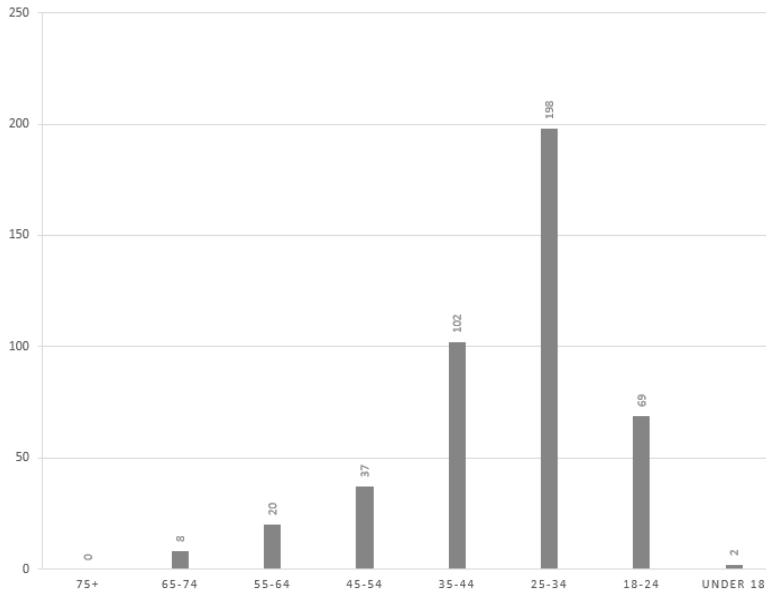
Figure 1. *Number of participants distributed by age demographic*

Furthermore, the sample demonstrated a robust spread across age categories according to gender (Figure 2). Although the majority of participants belonged to the 25-34 age bracket (46%), sufficient distribution across other age demographics existed to be representative. Moreover, our demographics reflect external statistics regarding average age of computer users in the U.S. (Anderson, 2015). An important note regarding the two participants self-reporting an age as *Under 18*: our IRB review included the potential for protected category participants. Given the anonymity of both our instrumentation and Mechanical Turk, the risk for harm to respondents was evaluated as minimal.

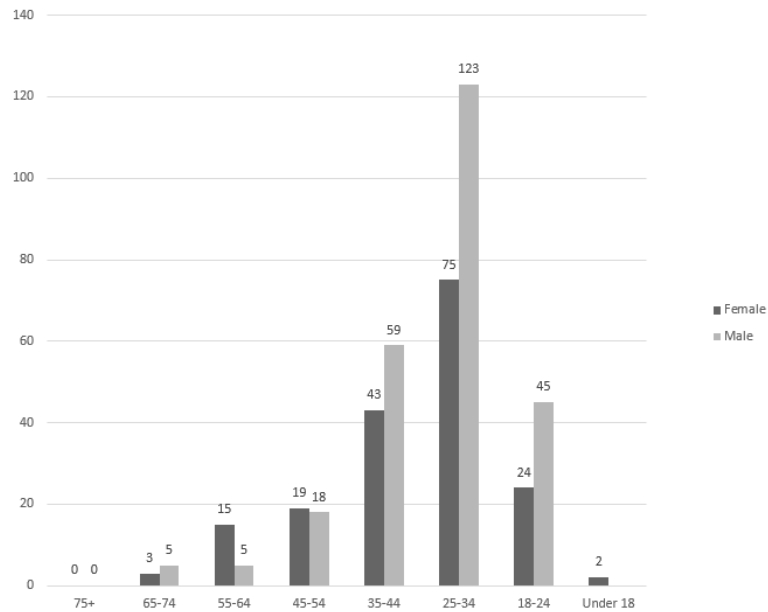
Figure 2. *Number of participants distributed by age and gender demographics*

Level of education was another socioeconomic characteristic we captured (Figure 3). We offered the nine categories for participants to offer education level. As shown in the table below, we found the 202 participants had received an *undergraduate degree*, which was double the amount who had received a *graduate level degree*. However, our sample showed a wide range of education, which helped to identify patterns related to password identification. It is interesting to note that the majority of males and females had received a *bachelor's degree*, with the second highest category being a *master's* or *graduate level degree*. There were also a higher number of males and females which had received *some college credits* while less had actually received an *Associates degree*. There were only 9 total participants who had received a Doctorate level degree.

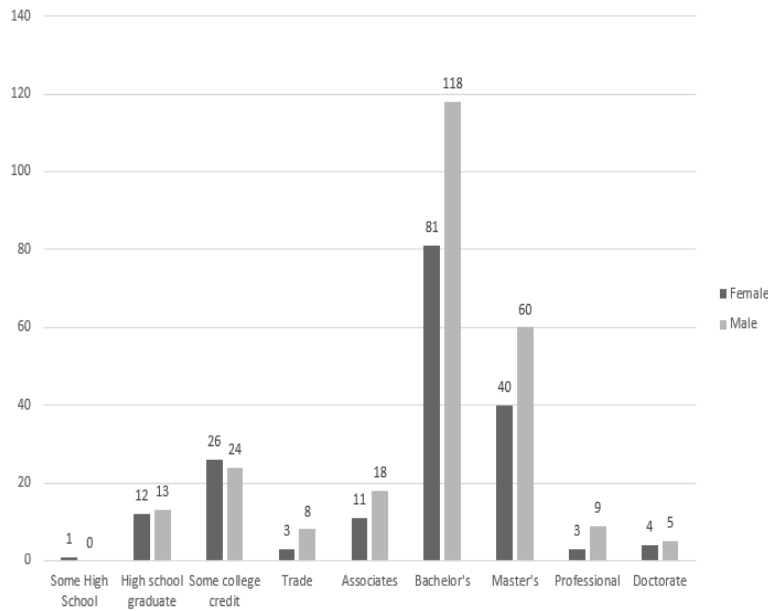

Figure 3. *Frequency of participant self-reported highest level of education completed by gender*

As well, we identified 33 profession categories based on similar demographic instrumentation in the literature (US census citation). While we did not limit participation by profession in any way, our aim was to collect a heterogeneous sample as to reasonably present a wide spectrum of *user* perception and ability. The top three professions for females were *Homemaker* (30), *Student* (17), and a tie for third with *Unemployed* (13) and *Education - College* (13). For males, the top three professions were *IT - Management* (53), *Student* (23), and *Telecommunication* (21).

**Table 1**. *Participant self-reported profession by gender*

| Profession | F | M | Profession | F | M |
|---|---|---|---|---|---|
| Homemaker | 30 | 2 | IT | 1 | 4 |
| Retired | 2 | 4 | IT - Management | 12 | 53 |
| Student | 17 | 23 | IT - Executive | 4 | 12 |
| Unemployed | 13 | 15 | Processing | 5 | 4 |
| Agriculture | 0 | 4 | Legal | 1 | 0 |
| Arts | 12 | 7 | Manufacturing - Electronics | 0 | 1 |

| Table 1 (continued) | | | | | |
|---|---|---|---|---|---|
| Broadcasting | 0 | 1 | Manufacturing - Other | 1 | 1 |
| Education - College | 13 | 20 | Military | 4 | 10 |
| Education - K-12 | 7 | 2 | Mining | 0 | 1 |
| Construction | 4 | 2 | Publishing | 0 | 1 |
| Finance | 4 | 8 | Real Estate | 3 | 3 |
| Government | 8 | 14 | Religious | 2 | 1 |
| Health Care | 3 | 9 | Retail | 1 | 0 |
| Hotel & Food Service | 11 | 7 | Telecommunication | 5 | 21 |
| Wholesale | 1 | 2 | Transportation | 4 | 6 |
| Scientific | 3 | 6 | Utilities | 3 | 8 |
| Software | 5 | 5 | | | |

A sample bereft of experience with passwords would not be ideal given our aim to measure potential relationships between socioeconomic attributes and identification of weak and strong passwords. Thus, we asked participants to report on the number of passwords used on a daily basis (Figure 4). We found a majority of participants used between one and five passwords daily (65%) whereas a stark minority used more than 11 passwords daily (5%).

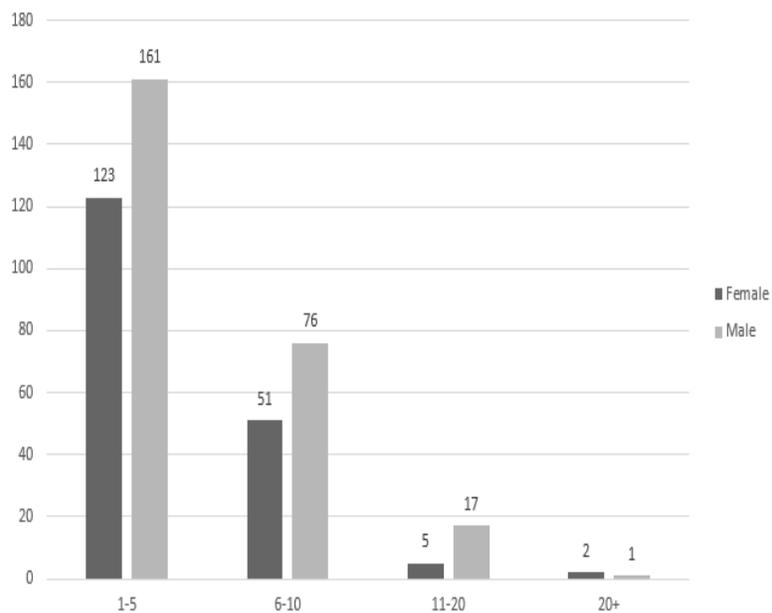

Figure 4. *Daily frequency of passwords used by participants by gender*

Lastly, participants self-reported a level of technical skill compared to people they knew (Figure 5). Combined males and females considered themselves *more skilled* with a total of 214 total participants. In comparison, 113 participants noted they had the *same skill* as the other people they were familiar with. It could be interesting in a later study to find out what subjective or objective metrics individuals use to determine they are more adept with technology than others around them.

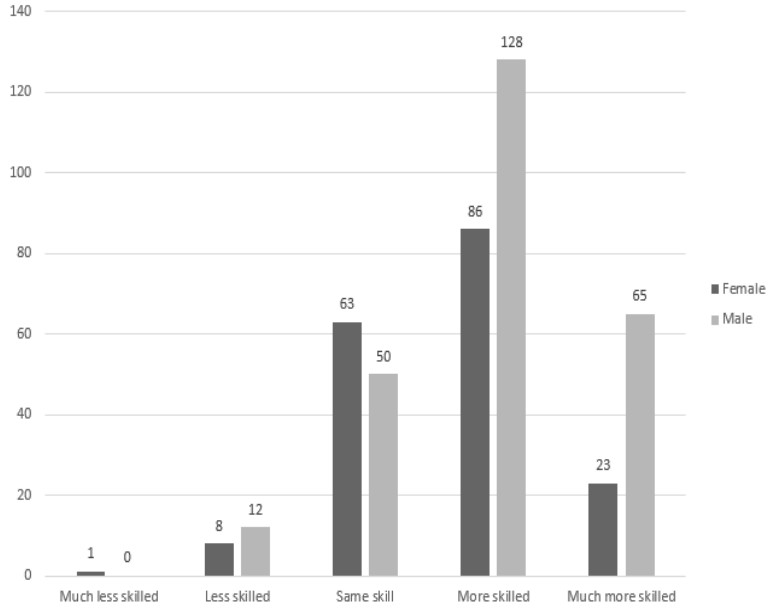

Figure 5. *Frequency of participant self-reported level of technical skill by gender*

*Hypotheses*

Initially, we broadly conjectured participant education, profession, and technical skill would show a relationship with successful identification of weak and strong passwords. However, accurate measurement of potential relationships would be cumbersome given the multi-level structure of our variables. Thus, we opted to disambiguate our general hypothesis into the following, more precise and testable constructs.

**Table 2**. *Testable hypotheses according to conjectured variable relationships*

| Hy | Relationships | Hypotheses |
|----|---------------|------------|
| 1 | *Level of education* and weak passwords | There is a relationship[1] |
| 2 | *Level of education* and strong passwords | There is a relationship[1] |
| 3 | *Profession* and weak passwords | There is a relationship[1] |
| 4 | *Profession* and strong passwords | There is a relationship[1] |
| 5 | *Technical skill level* and weak passwords | There is a relationship[1] |
| 6 | *Technical skill level* and strong passwords | There is a relationship[1] |

Note. The associated *null hypothesis* posits a relationship exists. The associated *null hypothesis* posits no relationship exists

**Results**

Would participant education, profession, or technical skill level be related to successful identification of weak and strong passwords? The answers we found were all affirmative except for one variable combination. The source for these answers was a sample group of 436 human participants who generated 21,800 discrete trials (number of participants [436] multiplied by the number of total password interactions per individual subject [50]).

We used a Chi-square test of independence to establish whether statistical relationships existed per our stated hypotheses with guidance from McDonald (2009). Further, we finished by analyzing the nominal strength of the most frequently identified and misidentified passwords according to *strength*. Here, we employed entropy analyses aligned with standardized entropy measures per NIST SP 800-63 (Greene et al., 2016). While not associated with our primary focus, we felt at least a cursory description using of these results may shed light on participants correctly or incorrectly identified passwords as weak and strong. We were aware of efforts to develop more precise strength measurements (Bonneau, 2012), however we opted to rely on published national standards because of the pervasive reliance of NIST specifications.

*Education*
We explored education as a potentially related variable first. Here, we hypothesized that education would demonstrate a relationship. To test the conjecture, we compared the level of education (9 levels) and frequency of correctly and incorrectly identifying both weak and strong passwords (Table 3).

**Table 3**. *Frequencies of participant level of education and identification of passwords*

| Education | Weak Correct | Weak Incorrect | Strong Correct | Strong Incorrect |
|---|---|---|---|---|
| Some High School | 14 | 11 | 13 | 12 |
| High School | 439 | 186 | 305 | 320 |
| Some College | 962 | 338 | 644 | 656 |
| Trade School | 257 | 68 | 149 | 176 |
| Associate's | 544 | 206 | 374 | 376 |
| Bachelor's | 3804 | 1246 | 2282 | 2768 |
| Master's | 2011 | 564 | 1268 | 1307 |
| Professional | 204 | 96 | 157 | 143 |
| Doctorate | 130 | 95 | 116 | 109 |

The Chi-square test of independence comparing education to successful identification of weak passwords (columns two and three in Table 3) revealed the variables to have a significant relationship ($X^2$ [8] = 64.89, *p* value of 5.10E-11  0.05 and a *critical value* of 15.5). As such, we rejected the associated null hypothesis (Table 2: *Hy 1*). Each group correctly identified weak passwords more frequently than incorrectly. The groups which were able to identify weak passwords 3 times more often than incorrectly were the individuals with bachelor's degrees.

For weak password analysis, participants with some high school and individuals at the doctorate level identified correct passwords 56% and 57% of the time, respectively. The other education levels, High School, Some College, Trade School, Associate's, Bachelor's, Master's, and a form of Professional degree, ranged from 68% to 79% able to correctly identify weak passwords. All education levels were able to correctly spot weak passwords 70% of the time. While it is interesting that the least amount of education (some high school), and most (Doctorate), had the lowest average, the overall average displayed the ability for participants to find the weak password.

The second step compared education to successful identification of strong passwords (columns four and five in Table 3). Here as well we found a significant relationship between variables ($X^2$ [8] = 33.71, *p* value of 4.58002E-05 0.05 and a *critical value* of 15.5). Thus, we rejected the null hypothesis (Table 2: *Hy 2*) as well.

Overall, all education levels were within an 8-point percentage variance, between 45 and 52%, able to spot a strong password. The groups with the highest scores of 52% were Some High School and Professional degrees, with a Doctorate level education missing by one point at 51%. Average ability to spot a strong password was at 49%, showing a major decline in ability to spot a strong password. Each group of education level was able to spot a weak password than a strong password more often.

**Table 4**. *Frequencies of participant profession and identification of passwords*

| Education | Wk. Correct | Wk. Incorrect | Str. Correct | Str. Incorrect |
|---|---|---|---|---|
| Homemaker | 654 | 46 | 402 | 498 |
| Retired | 101 | 6 | 78 | 115 |
| Student | 715 | 74 | 486 | 725 |
| Unemployed | 498 | 51 | 341 | 510 |
| Agriculture | 98 | 11 | 45 | 46 |
| Arts | 315 | 16 | 250 | 369 |
| Broadcasting | 10 | 0 | 14 | 26 |
| Education - College | 575 | 59 | 403 | 613 |
| Education - K-12 | 200 | 34 | 92 | 124 |
| Construction | 124 | 15 | 69 | 92 |
| Finance | 178 | 8 | 160 | 254 |
| Government | 412 | 50 | 258 | 380 |
| Health Care | 219 | 11 | 157 | 213 |
| Hotel & Food | 346 | 19 | 233 | 302 |
| Info. Technology | 114 | 6 | 64 | 66 |
| Info. Technology Mgmt | 1377 | 138 | 772 | 963 |
| Info. Technology Exec | 318 | 31 | 193 | 258 |
| Processing | 158 | 14 | 112 | 166 |
| Legal | 8 | 0 | 14 | 28 |
| Manufacturing - Electronics | 16 | 14 | 0 | 20 |
| Manufacturing - Other | 29 | 3 | 25 | 43 |
| Military | 261 | 30 | 166 | 243 |
| Mining | 27 | 0 | 14 | 9 |
| Publishing | 15 | 0 | 14 | 21 |
| Real Estate | 99 | 2 | 82 | 117 |
| Religious | 51 | 20 | 22 | 57 |
| Retail | 8 | 0 | 14 | 28 |
| Scientific | 162 | 26 | 100 | 162 |
| Software | 228 | 22 | 118 | 132 |
| Telecommunications | 605 | 57 | 307 | 331 |
| Transportation | 164 | 11 | 129 | 196 |
| Utilities | 217 | 17 | 137 | 179 |
| Wholesale | 63 | 5 | 37 | 45 |

*Profession*
The second variable we evaluated for a relationship was participants' self-reported profession. Here, again, we hypothesized that the variable would exhibit a relationship. To analyze the possible relationship, we examined the frequency at which 33 profession categories correctly or incorrectly identified weak and strong passwords.

Profession exhibited a significant relationship with identifying weak passwords ($X^2$ [32] = 153.19, *p* value of 0.00 0.05 and a *critical value* of 46.19). Consequently, we rejected the null hypothesis (Table 2: *Hy 3*).

The only professions to demonstrate less than 80% accuracy in identifying weak passwords were Manufacturing - Electrical (53%) and Religious (71%). The professions which were able to identify with between 80 and 89% were Agriculture, Education-K-12, Construction, Government, and Scientific. All other categories of professions were able to correctly find weak passwords over 90% of the time. Lastly, the professions which were able to correctly identify weak passwords 100% of the time were Broadcasting, Legal, Mining, Publishing, and Retail. Of all professions, the average score for finding weak passwords was 90%.

We also evaluated participants' profession against identification of strong passwords. We rejected the null hypothesis (Table 2: *Hy 4*) for this variable as well based on the Chi-square test of independence results ($X^2$ [32] = 66.11, *p* value of 0.0003 0.05 and a *critical value* of 46.19).

Strong passwords saw a much lower result from weak passwords, with an average of all professions only able to identify them 40% of the time. The professions with scores lower than 40% were Broadcasting, Education-College, Finance, Legal, Manufacturing - Other, Religious, and Transportation. This was interesting because both Broadcasting and Legal professions found the weak passwords 100% of the time. The only group which was not able to identify any strong passwords was Manufacturing - Electrical, which showed this group had the lowest identification for both weak and strong passwords combined. The group with the highest overall identification rate of 60% was Mining.

*Technical Skill*
We evaluated participant technical skill as the third and final variable. As before, we hypothesized technical skill, even self-reported technical skill, would exhibit a relationship with identifying weak and strong passwords. We examined the potential relationship according to four skill categories as follows.

**Table 5**. *Frequencies of participant technical skill and identification of passwords*

| Technical Skill | Wk. Correct | Wk. Incorrect | Str. Correct | Str. Incorrect |
|---|---|---|---|---|
| Much less skilled | 16 | 6 | 8 | 20 |
| Less skilled | 355 | 23 | 271 | 401 |
| Same skill | 2119 | 201 | 1367 | 1913 |
| More skill | 4097 | 411 | 2585 | 3607 |
| Much more skilled | 1778 | 155 | 1077 | 1390 |

The Chi-square test of independence comparing self-reported technical skill to successful identification of weak passwords demonstrated a significant relationship ($X^2$ [4] = 14.95, *p* value of 0.005 0.05 and a *critical value* of 9.5). As such, we rejected the associated null hypothesis (Table 2: *Hy 5*).

Overall it seemed that the participants were able to accurately depict technical skill level when it came to perceiving weak passwords. The group which self-reported as *much less skilled* than their counterparts averaged 72% when identifying weak passwords. The *less skilled* group up to the *much more skilled* group averaged 91% ability to identify weak passwords. Of all the groups, participants averaged an 87% ability to find weak passwords correctly. While significantly, this is a poorer performance compared to the relationship between profession and password strength perception.

Finally, we compared technical skill to successful identification of strong passwords. Unlike with previous tests, we found no relationship between these variables ($X^2$ [4] = 5.93, *p* value of 0.205 0.05 and a *critical value* of 9.5) and therefore accepted the null hypothesis (Table 2: *Hy 6*). This was a surprising result to say the least and warrants deeper investigation.

*Password inferences*

After completing the Chi-square tests for independence, we wanted to more closely examine details associated with passwords used in the instrument (Figure 6). The goal was to develop a richer inference of what participants perceived based on their *weak* or *strong* selections and the entropy of associated passwords. We offer only descriptive analysis here of associations between data points; there was no attempt to infer causation.

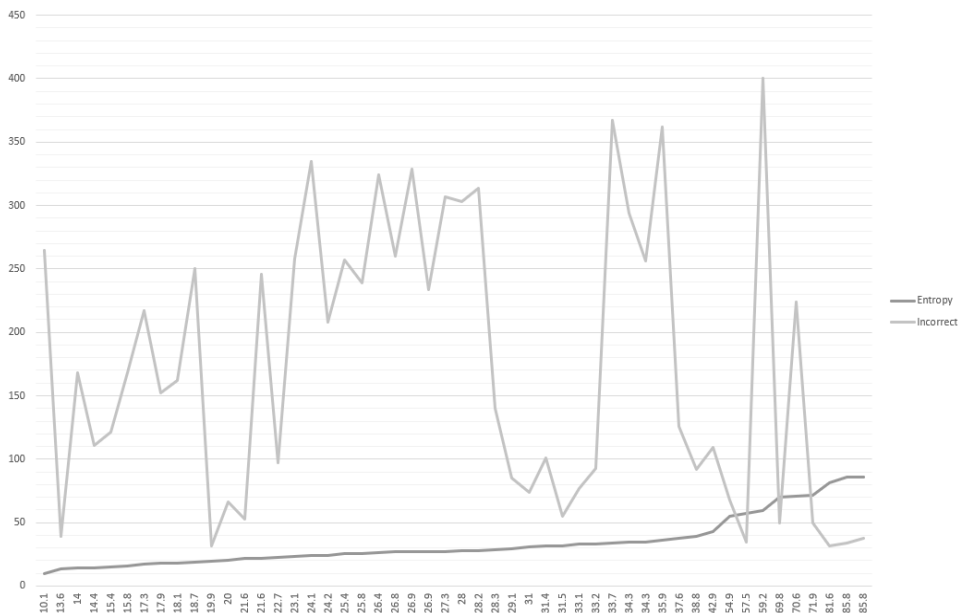

Figure 6. *Trend of incorrect perception of password strength as entropy in string rises*

To begin, the most common incorrect selection was for the string *G@m30f7hr0n3$*. Four hundred participants incorrectly perceived this to be a weak password compared to 36 that correctly identified this as a strong password. In reality, the entropy for the string is 59.2. Conversely, the most common correct selection was a tie between the string *1FcgiEF46Xy06jVS1* and *qwerty*. The former was a strong password whereas the latter was a weak password, and both were identified by 415 participants as such. The entropy of the two strings was 81.6 for the strong password and 19.9 for the weak.

The biggest misconception for users was about strong passwords; overall, participants had a harder time identifying strong passwords. It would be interesting to find out why one password *G@m30f7hr0n3$* was misidentified so frequently as a weak password. Further, if users felt this was such a weak password, why did they correctly identify *1FcgiEF46Xy06jVS1* as a strong password? We speculate that *leetspeak* may have signaled an equivalence to a corresponding plain text string.

**Conclusions**

The purpose of this work was to measure to what extent participant education, profession, or technical skill level are related to successful identification of weak and strong passwords. Towards this goal, we asked 436 human participants to judge whether 50 passwords were *weak* or *strong*. After data collection, we ran descriptive statistics so as to establish essential features and summaries for the sample. We then proceeded with a Chi-square test for independence to measure possible relationships between variables and evaluate our hypotheses which are summarized as follows (Table 6).

**Table 6**. *Summary of hypothesis testing results*

| Hy | Relationships | Results |
| --- | --- | --- |
| 1 | *Level of education* and weak passwords | There is a relationship |
| 2 | *Level of education* and strong passwords | There is a relationship |
| 3 | *Profession* and weak passwords | There is a relationship |
| 4 | *Profession* and strong passwords | There is a relationship |
| 5 | *Technical skill level* and weak passwords | There is a relationship |
| 6 | *Technical skill level* and strong passwords | There is no relationship |

Education was significantly related to successful identification of weak and strong passwords alike. Further, each individual educational strata demonstrated higher frequencies of correct identification than incorrect. Based on these results, we can infer perception of what constitutes a weak or strong password are not confined to any one educational stratum. While there could be benefit in more granular study here, we feel the results are robust enough to stand alone.

Profession was significantly related to successful identification of weak and strong passwords too. However, here we begin to see individual profession strata incorrectly identify password strength more often than correctly. While a stratum like *Homemaker* or *Retired* may not surprise anyone, the three *Information Technology* strata all more frequently misidentified strong passwords which is surprising to us. The inclusion of a variety of professions gave us enough of a range to see how each industry would perceive passwords. It is possible with an even larger set of professions to choose from, we could see a difference in password cognition.

There are a variety of follow up questions to be explored within the coupling of profession and perception of password strength. Experimental follow up may be of future interest to uncover what precisely causes specific professions to correctly identify weak passwords but incorrectly identify strong passwords. For a future study, we could focus specifically on individuals in IT fields, but target systems administrators, database administrators, and the like.

Interestingly, self-reported technical skill was significantly related to identifying weak passwords but not related to strong passwords. We are left to wonder about the potential underlying factors contributing to this situation and emphatically suggest follow up research in this area. Because we observed some trending towards incorrect identification of strong passwords in various professions (e.g., *Information Technology*), we must wonder if such professions inherently harbor mentalities associated with incorrectly identifying strong passwords. That stated, we suggest any future work with technical skill not rely on self-reporting. Such study could robustly establish technical skills through empirical instrumentation.

*Future work*
Future work on this topic could proceed in several directions. Based on our findings, we recommend additional qualitative study to provide further understanding of why individuals are not as capable of identifying strong passwords versus weak passwords. Interviewing participants with the aim of identifying how they determine password strength could give clarity to the results of this study. Furthermore, the results of this study may be useful in creating better training for users. Perhaps such training could adopt a gamified or games-based learning approach to reinforce patterns of relative password strength. Additionally, research should continue to probe *why* users use comparatively weak passwords despite evidence suggesting they perceive password strength. There exists a possibility that what users report as a perception of *weak* or *strong* does not match what they phenomenally experience when interacting with passwords. Here, brain-machine interface technology may be useful to design objective probing of what is inherently subjective.

Carnavalet, X. C. and Mannan, M. (2014). From Very Weak to Very Strong: Analyzing Password-Strength Meters. In: *NDSS '14*. San Diego: Internet Society, Available at: https://www.ndss-symposium.org/ndss2014/programme/very-weak-very-strong-analyzing-password-strength-meters/ [accessed August 24, 2019]

De Joode, D. (2012). *Does Password Fatigue Increase the Risk on a Phishing Attack?* Master's. Tilburg University.

Dhamija, R., and Dusseault, L. (2008). The Seven Flaws of Identity Management: Usability and Security Challenges. *IEEE Security and Privacy*, 6(2), pp. 24-29. doi: 10.1109/MSP.2008.49

Florencio, D., and Herley, C. (2007). A Large-Scale Study of Web Password Habits. In: *Proceedings of the 16th International Conference on World Wide Web*, Banff: ACM, pp. 657-666. doi: 10.1145/1242572.1242661

Grassi, P., Garcia, M., and Fenton, J. (2017). *Special Publication (SP) 800-63C: Digital Identity Guidelines*. National Institute of Standards and Technology, *doi: 10.6028/NIST.SP.800-63c*

Greene, K. K., Franklin, J. M., Greene, K. K., and Kelsey, J. (2016). *Measuring the Usability and Security of Permuted Passwords on Mobile Platforms*. Special Report 8040, US Department of Commerce, National Institute of Standards and Technology.

Hart, D. (2015). Two Studies on Password Memorability and Perception, In: *Proceedings of the 10th Annual Symposium on Information Assurance (ASIA '15)*, Albany: ASIA, pp. 7-13. Available at: https://www.albany.edu/iasymposium/proceedings/2015/ [accessed August 24, 2019]

Jansen, W. A. (2003) Authenticating Users on Handheld Devices. In: *Proceedings of the Canadian Information Technology Security Symposium.* Available at: https://www.nist.gov/publications/authenticating-users-handheld-devices [accessed August 24, 2019]

Jermyn, I., Mayer, A., Monrose, F., Reiter,, M., and Rubin, A. (1999). The Design and Analysis of Graphical Passwords. In: *Proceedings of the 8th Conference on USENIX Security Symposium*. Washington, DC: USENIX Association, p. 1. Available at: https://pdfs.semanticscholar.org/0201/bb5dc0345477d91f7192cea2a3230c0c638a.pdf [accessed August 23, 2019]

Komanduri, S., Shay, R., Cranor, L. F., Herley, C., and Schechter, S. (2014). Telepathwords: Preventing Weak Passwords by Reading Users' Minds. In: *Proceedings of the 23rd USENIX Conference on Security Symposium*. San Diego: USENIX Association, pp. 591-606. Available at: http://cups.cs.cmu.edu/rshay/pubs/telepath.pdf [accessed August 23, 2019]